# Scene-based nonuniformity correction with homography transformation

Peretz Yafin,[1,2], Nir Sochen,[3] and Iftach Klapp[1*,]

[1]*Department of Sensing, Information and Mechanization Engineering, Volcani Institute-ARO, Bet Dagan, Israel.*
[2]*School of Computer Science, Tel Aviv University, Tel Aviv 69978, Israel.*
[3]*Department of Applied Mathematics, Tel Aviv University, Tel Aviv 69978, Israel.*

**Corresponding author: iftach@volcani.agri.gov.il*

**Abstract:**

Due to their affordable, low mass, and small dimensions, uncooled microbolometer-based thermal focal plane arrays (UC-FPAs) are useful for long-wave infrared (LWIR) -imaging applications. However, in outdoor conditions typical in agricultural remote sensing, cameras based on UC-FPAs may suffer from drift in offset and gain. To tackle the persistent drift, the system requires continuous calibration. Our goal in this study was to eliminate this requirement via a computational schema. In a former study, we estimated unknown gain and offset values and thermographic images of an object from a sequence of pairs of successive images taken at two different blur levels. In the current work, we took on a similar problem using a sequence of shifted images, with relative shifts caused by realistic drone hovering modeled by homography transformation. This places our work in the realm of scene-based nonuniformity correction problems. We show that an object's thermographic values, as well as gain and offset, can be jointly estimated by relying on a few sets of shifted images. We use a minimum likelihood estimator, which is found using alternating minimization. Registration is done using the generalized Lucas–Kanade method. Simulations show promising accuracy with mean Pearson correlation of more than 0.9999998 between ground truth and restoration. Under ideal assumptions, this is equivalent to a mean restoration error of less than 0.01°C.

*Index Terms*—Imaging, Inverse problems, Functional analysis, Blind deconvolution.

## 1. Introduction

The emergence of the uncooled microbolometer-based (MB) thermal focal plane arrays (UC-FPA) has opened up a new world of applications for infrared detection [1]. Sensors of this type have improved features, such as small size, low power consumption, and affordability. Consequently, these cameras are used in various military and commercial applications [1], and are the most commonly used cameras in thermography [2]. The UC-FPA can be mounted on microdrones due to its compactness. This arrangement is valuable for scanning large areas, as in, for example, agricultural and environmental monitoring. In a recent work, Korsaeth and Kusnierek [3] used this arrangement to acquire a thermographic image of crops. Nevertheless, even though a protective shield was mounted on the camera, wind velocity and changes in radiation level influenced the FPA temperature. Since convection from a surface is affected by fluid velocity [4], changes in FPA temperature might be explained by variations in wind velocity around the camera through the effect of the envelope temperature [2]. Body temperature is also influenced by energy balance with the surrounding [4]. The FPA's drift restricts the ability to extract thermographic information from the LWIR camera. Therefore, calibrating the camera's readout is a crucial aim.

Regular MB structure includes absorbing layer and a thermometer. The absorbing layer is warmed by LWIR radiation, the resistance of the thermometer is changed by its temperature. The MB's readout is in units of voltage or current [2], in cameras, it is converted to gray values (gv).

Every pixel the FPA is a MB each subjected to a different gain (g) and offset (d) such that the pixel's readout is:

$$\mathrm{y}(k,l)=\mathrm{g}(k,l)v(k,l)+\mathrm{d}(k,l)+\mathrm{n}(k,l) \quad (1)$$

where y(k,l) , g(k,l), d(k,l) ,and n(k,l) are the raw value, gain, offset and additive noise of pixel (k,l) respectively.
The additive noise in the pixel is assumed to be a random variable taken from a normal distribution $n \in N(0, \sigma^2)$ with zero expectation with standard deviation σ. The correct radiation value at that pixel is denoted by v(k,l). The pixels of the UC-FPA are not perfectly isolated, thus $g = g(T)$ and $d = d(T)$. While the gain may assume linearly depended on temperature *T,* the offset's dependence on *T* is a power series of *T* and may include derivatives of *T* [5]. In one specific condition, authors showed that offset may be characterized as a quadratic function of *T* [6]. Including the dependency on the temperature, the pixel's readout is:

$$\mathrm{y}(k,l,T)=\mathrm{g}(k,l,T)v(k,l,T)+\mathrm{d}(k,l,T)+\mathrm{n}(k,l) \quad (2)$$

In our previous work [7], we suggested performing nonuniformity correction (NUC) using a sequence of pairs of images, each pair including a sharp image and a defocused one. The defocused image gave redundancy, making the problem solvable. In this work, we used small shifts as the source for redundancy. This makes the physical system less complex to build and avoids defocusing errors.
Our suggested method solves the problem of scene-based NUC. In this method, the input is a sequence of LWIR images corrupted by unknown gain and offset, from which a clean sequence of images with the correct value is extracted.
The thermal time constant is on the order of a minute while the imaging integration time is on the order of 10 ms [7] In this work, we exploit the relatively long time constant to assume that a quick sequence of images can be measured under the same gain and offset values [7].
Image registration has been used to perform NUC in several papers. Hardie et al. [8] used subpixel-shifted images to estimate gain and offset, while assuming a Gaussian pattern of offset and a global shift of all pixels. Similar assumptions of a global shift were assumed by Ratliff et al. [9]. In another work, Meza et al. [10] used an alternating gradient descent with a regularization term to perform NUC for a sequence of rotated and translated images, and Hardie and Droege [11] used maximum a posteriori estimator (MAP) to perform NUC with a sequence of rotated and translated images. Liu et al. [12] used small shifts between images to compensate for the offset without actually computing it. The paper most closely related to the present work is by Zhao and Zhang [13], who assumed a parametric motion model of affine transformations. They formulated the correcting problem as nonlinear least squares, which they solved by alternating minimization. However, their algorithm only succeeded in correcting the offset.

This paper shows how to correct both unknown gain and offset of a drifted LWIR camera, using a sequence of shifted images taken under different unknown homography transformations. We offer that this can be done by repeating the registration step during the alternating algorithm, which drastically improves the registration's precision and consequently improves the image's estimation. As it is capable of using images subjected to homography transformations, we allow our algorithm to handle realistic sequences of images that were taken by an unmanned aerial vehicle (UAV). The homography transformations take into account the UAV's vibrations, including translations, rotations, and tilts. In addition, using simulated shifted images, we show that our algorithm can correct both the gain and the offset, rather than only offset as previously done [10].

## 2. NUC with homography transformations

### 2.1. Image formation

In the previous section, we elaborated on the relation between the radiance that arrives to the bolometer and its electronic readout that is eventually depicted as a grey level of a pixel in the image. In this section, we add another layer which is the transformation of the temperature of a point on the ground in the field of view of the camera, denoted $x(k,l)$, to the radiance that is captured in each bolometer in the camera's array, denoted $v(k,l)$. The process is well approximated by geometric optic arguments and is pictorially described in Fig. 1. The optical process is defined by the Point Spread Function (PSF), which acts spatially on the radiance field by a convolution. Discretizing and working on the image in a column stack enables to an expression of the convolution operator as matrix multiplication. We denote the PSF by a matrix $A$ and denote the radiance that arrives at the $(k,l)$ bolometer by $v(k,l) = (Ax)(k,l)$. The equation for the image formation of a single image is:

$$y(k,l) = g(k,l)(Ax)(k,l) + d(k,l) + n(k,l) \quad (3)$$

We aim to extract the temperatures associated with radiation $x(k,l)$ from the y(k,l) readouts. This is, of course, an ill-posed problem. The drone takes many pictures while scanning or hovering over a field cell in practice. Since the images are taken at intervals of microseconds and the temperature change of the ground and of the drone is of an order of seconds, we have points on the grounds that appear in many different images. This redundancy is exploited here to make it well-posed and solve for the temperature at the ground without the need to calibrate the camera. Before we can write the set of equations to solve, we need to address the issue of the angle between the camera and the ground. The angle between the drone and the ground constantly changes either because of topographic change or because of small rolls, pitches and yaws of the drone. These changes of angles between images must be considered in the registration of the images so that we can correctly identify the same point on the ground in many different images. In general, the transformation between images is projective, but if the drone is high enough (a couple of dozen meters), it is well approximated by a homographic transformation that takes planes to planes. Suppose that we have two images $y_i$ and $y_j$. All the points $x(k,l)$ that appear in both images' fields of view are related by a homographic transformation. Transforming the point $(k,l)$ in $y_i$ to the homogenous coordinate system $(k,l,1)$ we can describe the action of the homographic transformation of $y_i$ to $y_j$ by a matrix that we denote by $H_{ij}$. The action on the whole column stack is described by the matrix $S^{Hij}$. We are finally ready to write the relation between each point $x_i(k,l)$ on the ground that was registered by image $y_i$ to the readout of any other image in which it appears

$$y_j(k,l) = g(k,l)(AS^{Hij}x_i)(k,l) + d(k,l) + n(k,l) \quad (4)$$

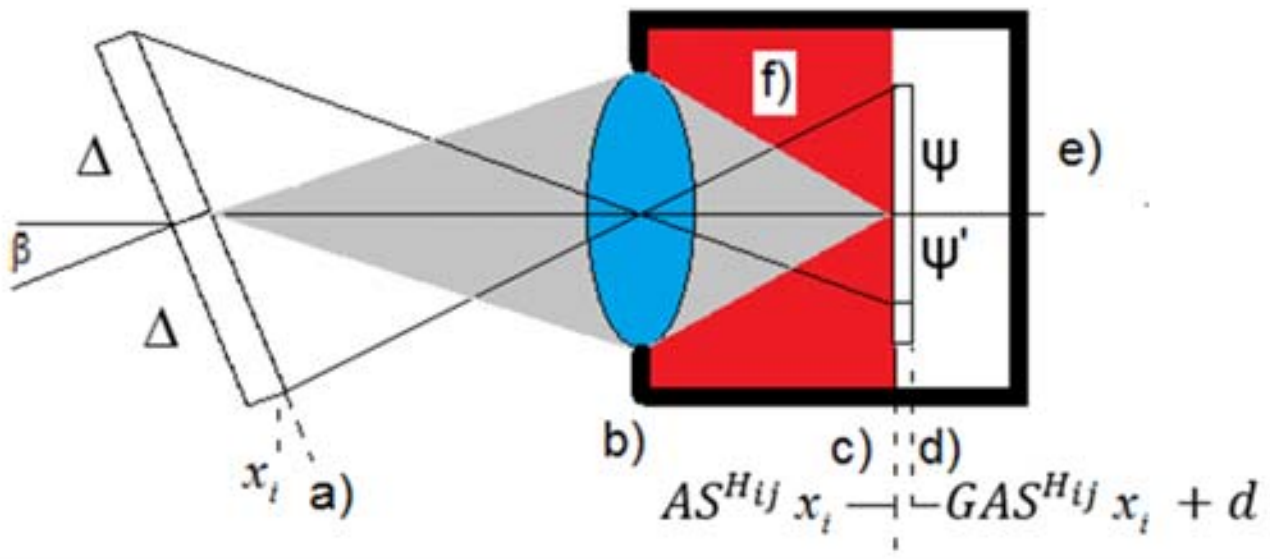

Fig. 1. Image formation

In column stack notation $y_j$, $g$, $d$, and $x_j$ are vectors. Denote by $G = \text{diag}\{g\}$ the diagonal matrix where the main diagonal is the vector $g$. The equation is written now as:

$$y_j = GAS^{Hij}x_i + d + n \quad (5)$$

Note that if we are given the exact homographies $H_{i,j}$ between each pair of images and the true radiation $x_i$, the problem becomes a regression problem for $g$ and $d$. Inversely if $H_{i,j}$, $g$ and $d$ are known then finding $x_i$ is a standard mean squares problem. Finally, if $g$, $d$ and $x_i$ are known we are in a problem which resembles a registration problem. In a regular registration problem we try to find the transformation that matches the readouts $y_i$s. In our case, because we have an image formation problem, we try to find the transformation that matches the actual temperatures $x_i$s. In the registration step one should take care to use for each pair $(i,j)$ of images only those points on the ground $x(k,l)$ that are captured in both images. In order to take that in account we define the filter $w_{ij}(k,l)$ that is equals 1 if the point on the ground $p = (k,l)$ appears in both images $y_i$ and $y_j$ and it is 0 otherwise. In a column stack arrangement $w_{ij}$ is a vector and we define the matrix $W_{ij} = \text{diag}\{w_{ij}\}$. To proceed we describe the problem as a Maximum Likelihood (ML) problem that will be solved by alternating between registration-like step, regression step and mean squares step. The maximum likelihood is based on the assumption that the noise at each pixel is normal, namely that:

$$P(n(k,l)) = \frac{1}{\sqrt{2\pi\sigma^2}} \exp\left(-\frac{n(k,l)^2}{2\sigma^2}\right) \quad (6)$$

For the whole image, assuming independence of the noise between different sensors, we find that:

$$P(x, g, d, H_{ij} \mid \{y_j\}) = \; = \frac{1}{\sqrt{2\pi\sigma^2}} \exp\left(-\frac{1}{2\sigma^2} \sum_{i,j=1}^{m} \left\| W_{ij}\left(GAS^{Hij}x_i + d - y_j\right)\right\|^2\right) \quad (7)$$

Where ($m$) is the number of images. Before we give details of the different optimization steps we note on a global ambiguity.

### 2.2. Scale and shift ambiguity

For every solution $\{\hat{x}\},\hat{g},\hat{d},\{\hat{H}_{ij}\}$ for Eq. (5), let a, b ∈ R, and notice that:

$$\begin{aligned}
&\sum_{i,j=1}^{m}\left\| W_{ij}\left(\frac{1}{a}\hat{G}AS^{H_{ij}}(a\hat{x}_i+b)+\hat{d}-\frac{b}{a}\hat{g}-y_j\right)\right\|^2 \\
&=\sum_{i,j=1}^{m}\left\| W_{ij}\left(\frac{1}{a}\hat{G}AS^{H_{ij}}a\hat{x}_i+\frac{1}{a}\hat{G}AS^{H_{ij}}b+\hat{d}-\frac{b}{a}\hat{g}-y_j\right)\right\|^2 \\
&=\sum_{i,j=1}^{m}\left\| W_{ij}\left(\hat{G}AS^{H_{ij}}\hat{x}_i+\frac{b}{a}\hat{g}+\hat{d}-\frac{b}{a}\hat{g}-y_j\right)\right\|^2 \\
&=\sum_{i,j=1}^{m}\left\| W_{ij}\left(\hat{G}AS^{H_{ij}}W_j\hat{x}_i+\hat{d}-y_j{}^2\right)\right\|^2
\end{aligned} \tag{8}$$

*Hence,* $\left((a\hat{x}+b),\frac{1}{a}\hat{g},\hat{d}-\frac{b}{a}\hat{g}\right)$ *is also a solution for* Eq.7. Therefore, this restoration problem can be solved only up to a global scale and shift. In the simulations below, we handle this by using normalized simulated gain and offset, and the algorithm's results are constrained such that the gain mean is 1, and the offset mean is 0. This can be formally integrated in the formalism either by a prior and carrying a MAP approach or by a Lagrange multiplier approach. In practice we did after each step of an unconstrained optimization step a mapping into the space of allowed (g,d) pairs. These assumptions can be realistic if the initial correction of the system for a specific temperature is done by two-point correction [14]. In this case, the algorithm's results, after normalization, are approximately correct. Otherwise, the algorithm's restoration holds only up to a global scale and shift. We evaluate, therefore, our results using, inter alia, the Pearson correlation coefficient. Pearson's coefficient is a proper metric for measuring the quality of our algorithm since it is independent of scale and shift.

### 2.3. Algorithm

Similar to Zhao and Zhang [13], we solve Eq. 7. using alternate minimization, where the objective is minimized over $x$ and then over $g$ and $d$, repeatedly. It is natural to use alternate minimization in this problem that way since it becomes linearly related to $x$, ($g$ or $d$) when the two other variables are fixed, hence the problem can be solved efficiently at each stage using a linear solver or an explicit expression. However, in our algorithm, we continue to perform the registration stage during the alternating steps of the algorithm, rather than only before as in Zhao and Zhang [13]. The algorithm ends up with a mutual estimation of the series of images {x}, the unknown gain $g$, the unknown offset *d*, the registration, and the homography parameters. The resulting process is shown in Fig.2. A detailed account of the algorithmic process is given in sub-sections 1 to 6.

First we use the current estimates of *x*, *g,* and *d* to do a registration-like step and find the homography $H_{ij}^{n}$ by:

$$\begin{aligned} H_{ij}^{n} &= argmax_{H_{ij}} P\left(H_{ij} \mid \left\{x_i^n, g^n, d^n\right\}, \{y_j\}\right) \\ &= argmin_{H_{ij}} (-lnP\left(H_{ij} \mid \left\{x_i^n, g^n, d^n\right\}, \{y_j\}\right) \end{aligned} \quad (9)$$

In practice, we approximate this minimization, see step 6 below. Next, we use the homography $H_{ij}^{n}$ and the values $x_i^n$ to perform regression on *g* and *d.*

$$\begin{aligned} \left\{g^{n+1}, d^{n+1}\right\} &= argmax_{g,d} P\left(g, d \mid H_{ij}^{n}, \left\{x_i^n\right\}, \left\{y_j\right\}\right) \\ &= argmin_{g,d} (-lnP\left(g, d \mid H_{ij}^{n}, \left\{x_i^n\right\}, \left\{y_j\right\}\right) \end{aligned} \quad (10)$$

Finally, $x_i^{n+1}$ are estimated by the least squares method

$$\begin{aligned} \left\{x_i^{n+1}\right\} &= argmax_{x_i} P\left(x_i \mid g^{n+1}, d^{n+1}, H_{ij}^{n}, \left\{y_j\right\}\right) \\ &= argmin_{x_i} (-lnP\left(x_i \mid g^{n+1}, d^{n+1}, H_{ij}^{n}, \left\{y_j\right\}\right) \end{aligned} \quad (11)$$

The details are given in the next section.

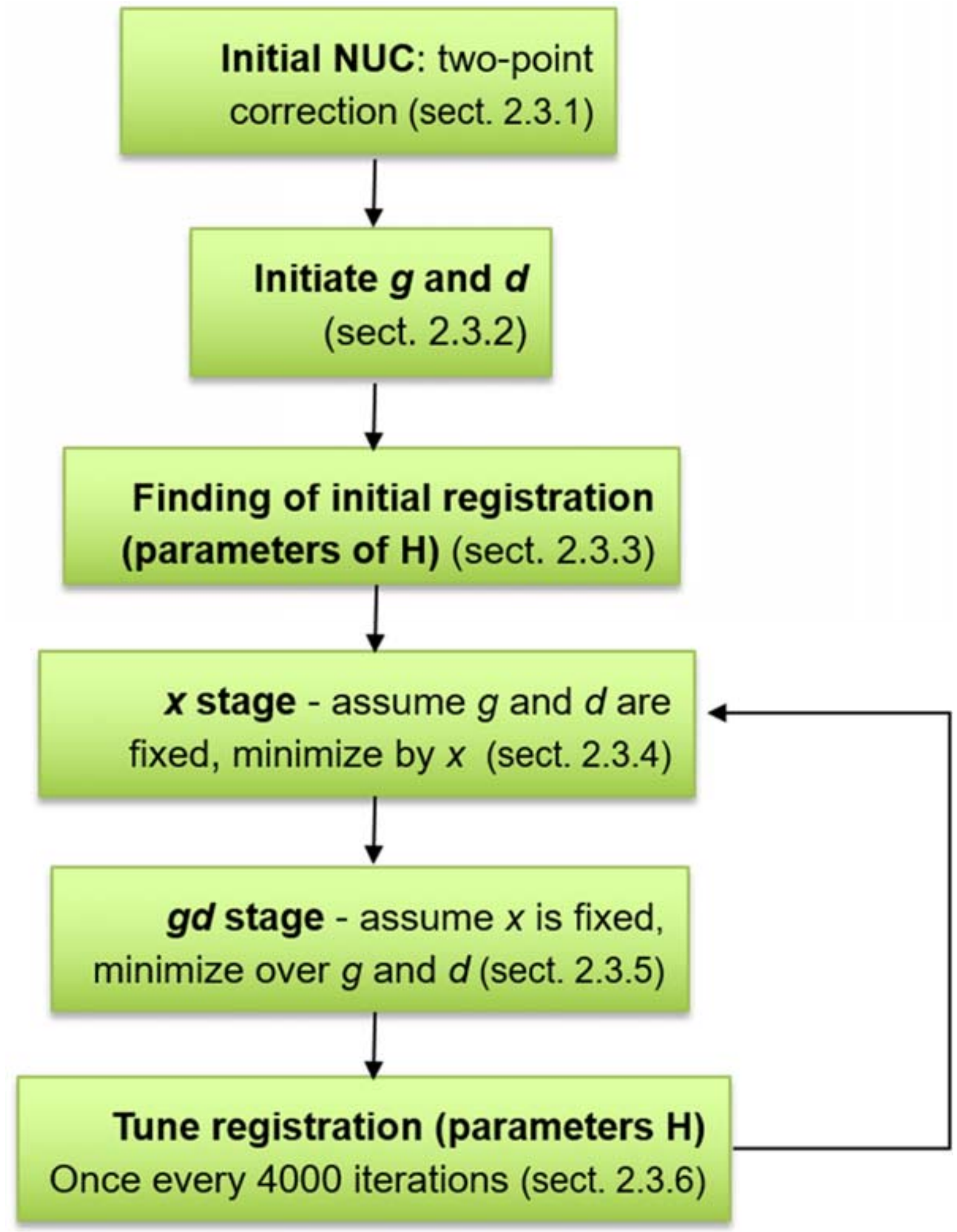

Fig. 2. Proposed algorithm. First, an initial NUC and rough estimation of g and d is performed, and then the initial registration is performed using the estimated g and d. All of the parameters then estimated in an alternating process.

### 2.3.1.Initial NUC

Similar to our previous work [7], it is assumed that the initial NUC is done by a two-point correction [14]. This stage is performed in a laboratory before the scene is imaged. It makes use of at least two different known blackbody temperatures. The blackbody is placed such that the surface temperature is uniformly distributed over the camera's entire field of view (FOV). The camera readout is taken, and it is linearly related in every pixel. For every pixel, slope (gain) and intercept (offset) are calculated directly or by least squares if more than two points are used. This stage provides an initial NUC for a specific environmental temperature. Assuming that gain and offset do not differ extremely at other temperatures, the initial estimation will be close to the real gain and offset.

### 2.3.2. Initiating g and d

The initial NUC (sect. 1) is performed once before the iterative phase, and is critical for achieving sufficiently accurate registration for the algorithm to converge. We apply, similar to Zhao and Zhang [13], a statistics-based NUC before the registration stage. As rough estimations for g and d, we use the standard deviation and mean, respectively, of all images of all ROIs:

$$d_0(p) = \frac{1}{n_p}\sum_{j=1}^{m} W_{i_0 j} y_j \qquad g_0(p) = \sqrt{\frac{1}{n_p}\sum_{j=1}^{m} W_{i_0 j}\left(y_j - d_0\right)^2} \qquad (12)$$

Where $i_0$ is the pivot object (for details see step 6 below), and $n_p = n(k,l)$ be the number of pair of images in which pixel $p$ appears in both images, namely $n_p = \sum_{i,j} W_{i,j}$ .

### 2.3.3. Initial registration

Initial registration is performed using the Lucas–Kanade method, as detailed in sect. 6. $H$ is initialized such that it expresses global shift:

$$H_{ij}^{0} = \begin{pmatrix} 1 & 0 & t_x \\ 0 & 1 & t_y \\ 0 & 0 & 1 \end{pmatrix} \qquad (13)$$

where the shift is found using simple template matching. Superscript denotes the zero iteration.

### 2.3.4. x stage

Fixing $\hat{d}$ and $\hat{g}$ and denoting $\hat{G} = diag(\hat{g})$, we get:

$$\hat{x} = argmin_x \sum_{i,j=1}^{m} \left\| W_{ij}\left(\hat{G}AS^{H_{ij}} x_i + \hat{d} - y_j\right) \right\|^2 \qquad (14)$$

For $x_i$ this is equivalent to the linear least squares solution of a block diagonal matrix, where the ($i^{th}$) block matrix equals:

$$\begin{pmatrix} W_{i1}\hat{G}AS^{H_{i1}} & \cdots & 0 \\ \vdots & \ddots & \vdots \\ 0 & \cdots & W_{im}\hat{G}AS^{H_{im}} \end{pmatrix} \begin{pmatrix} x_i \\ \vdots \\ x_i \end{pmatrix} = \begin{pmatrix} y_1 - \hat{d} \\ \vdots \\ y_m - \hat{d} \end{pmatrix} \qquad (15)$$

This is solved using LSQR [15] which is a matrix free method , i.e., there is no need to build a large matrix explicitly, but its transformation is applied and its transformation transposed, which can be done efficiently using fast implementation of convolution and image warping. The iteration number of the LSQR is limited to 20. In each new x stage the initial guess of the LSQR is the current $\hat{x}$ found by the previous LSQR.

### 2.3.5. g and d stage

Fixing x we get:

$$\hat{g},\hat{d} = argmin_{g,d} \sum_{i,j=1}^{m} \left\| W_{ij}\left( GAS^{Hij} x_i + d - y_j \right) \right\|^2$$
$$= argmin_{g,d} \sum_{p=1}^{hw} \sum_{i,j=1}^{m} W_{ij}\left( g(AS^{H_{ij}} \hat{x}_i) + d - y_j \right)^2 (p) \quad (16)$$

where the index $p$ goes over all of the (k,l) pixels in the image. Every summand depends only on g(p), d(p), hence the problem is separable pixel-wise and is equivalent to:

$$g(p), d(p) = argmin_{g(p),d(p)} \sum_{i,j=1}^{m} W_{ij}\left( g(AS^{H_{ij}} \hat{x}_i) + d - y_j \right)^2 (p) \quad (17)$$

for every pixel $p$. This is a simple linear regression problem, which has an explicit solution.

Let the mean of a tensor. Then the variance is $\langle A \rangle = \left(1/n_p\right) \cdot \sum_{i,j} W_{i,j} A_{i,j}$ be $A_{i,j}$

$$Var\left(A_{ij}\right) = \frac{1}{n_p} \sum_{i,j} W_{ij}\left(A_{ij} - \langle A \rangle\right)^2 \quad (18)$$

and the covariance is

$$COV\left(A_{ij}, B_{ij}\right) = \frac{1}{n_p} \sum_{i,j} W_{ij}\left(A_{ij} - \langle A \rangle\right)\left(B_{ij} - \langle B \rangle\right) \quad (19)$$

Then vector $\hat{g}$ is given explicitly by a pixel-wise division

$$\hat{g} = \frac{COV\left( AS^{H_{ij}}{}_j x_i, W_{ij} y_j \right)}{VAR\left( AS^{H_{ij}} x_i \right)} \quad (20)$$

and

$$\hat{d} = \left\langle W_{ij} y_j \right\rangle - \hat{g}\left\langle AS^{H_{ij}} x_j \right\rangle \quad (21)$$

### 2.3.6. Registration step

In order to reduce computation time, we choose pivotal images $x_{i_k}$ such that these pivotal images have no pixels in common, namely for any two pivotal images $x_{i_k}$ and $x_{i_l}$ the matrix $W_{i_k i_l} = 0$. Next let the pivotal image be w.l.o.g. $x_{i_0}$ and let $x_j$ be an image such that Wi0j is not identically zero. Define $\tilde{x}_j = (y_j - \hat{d})/\hat{g}$, then the homography $H_{i_0 j}$ is computed as:

$$\hat{H}_{i_0 j} = argmin_H \left\| S^H \tilde{x}_{i_0} - \tilde{x}_j \right\| \quad (22)$$

The $\tilde{x}_j$s are approximately "clean" of nonuniformity, and they can be registered using iterating generalized Lucas–Kanade, similarly to what was done in Capel and Zisserman [16].

Eq. (22). is solved by the Gauss-Newton method. This is performed for 50 iterations, where the initialization of $\widehat{H}_{i,j}$ by its previous estimation. For completeness, we now show the detailed derivation of the Gauss-Newton step.

Let $X_i: R^2 \to R$ be a continuous version of $\tilde{x}_i$, we want to register it with a shifted image $X_j: R^2 \to R$ a continuous version of $\tilde{x}_j$.

Let the function $e: R^2 \to R^3$ be such that $e = (s\ t\ 1)^T$. The image transformed by the homography matrix $H \in R^{3\times3}$ is:

$$\tilde{X}_i = \tilde{X}_i(s,t) = \tilde{X}_i(\tilde{s},\tilde{t}) = \tilde{X}_i\left(\frac{H_1 \cdot e}{H_3 \cdot e}, \frac{H_2 \cdot e}{H_3 \cdot e}\right) \quad (23)$$

where $H_i$ are the different rows of matrix $H$. According to the chain rule, the gradient with respect to $H$ is:

$$\frac{\partial \tilde{X}_i(\Phi, H)}{\partial H} = \nabla \tilde{X}_i \cdot \begin{pmatrix} \frac{e^T}{H_3 e} & 0 & -\frac{H_1 e}{H_3 e^2} e^T \\ 0 & \frac{e^T}{H_3 e} & -\frac{H_2 e}{H_3 e^2} e^T \end{pmatrix} =$$

$$\left( (\tilde{X}_i)_s \quad (\tilde{X}_i)_t \right) \begin{pmatrix} \frac{e^T}{H_3 e} & 0 & -\frac{H_1 e}{H_3 e^2} e^T \\ 0 & \frac{e^T}{H_3 e} & -\frac{H_2 e}{H_3 e^2} e^T \end{pmatrix} = \quad (24)$$

$$= \left( \frac{(\tilde{X}_i)_{\tilde{s}}}{H_3 e} \quad \frac{(\tilde{X}_i)_{\tilde{t}}}{H_3 e} \quad -(\tilde{X}_i)_{\tilde{s}} \frac{H_1 e}{H_3 e^2} - (\tilde{X}_i)_{\tilde{t}} \frac{H_2 e}{H_3 e^2} \right) \cdot e^T$$

Where $\Phi = (s,t), \tilde{\Phi} = (\tilde{s},\tilde{t})$ ,. $(\tilde{X}_i)_{\tilde{s}} = \frac{\partial \tilde{X}_i(\Phi)}{\partial \tilde{s}}, (\tilde{X}_i)_{\tilde{t}} = \frac{\partial \tilde{X}_i(\Phi)}{\partial \tilde{t}}$ ,and $(\tilde{X}_i)_s = \frac{\partial \tilde{X}_i(\Phi)}{\partial s}, (\tilde{X}_i)_t = \frac{\partial \tilde{X}_i(\Phi)}{\partial t}$

Hence, given the image points $(s_i, t_i)$ (in practice, points in the image boundaries are not taken), the Gauss–Newton step is:

$$H_{m+1} = H_m + \left(M^T M\right)^{-1} M^T \left(S^{H_{ij}} X_i - X_j\right) \quad (25)$$

where $M \in R^{hw\times9}$ is the Jacobian matrix $M^T M$ is a $9 \times 9$ matrix which can be inverted efficiently. The image derivatives $I_x$ and $I_y$ are computed numerically by applying on $\tilde{x}_{i,j}$ the filter $\left[-\frac{1}{2}\ 0\ \frac{1}{2}\right]$ in each axis. In every registration, $H$ is initialized with its last estimation from the previous registration stage. $M$ is calculated from the data, the expression for the $i^{\text{th}}$ row of $M^T$ is give in equation (19).

$$M_i^T = \frac{\partial \tilde{I}\left(s_i, t_i, H\right)}{\partial H} = \frac{1}{H_3 e}\begin{pmatrix} s_i(\tilde{X}_i)_{\tilde{s}} \\ t_i(\tilde{X}_i)_{\tilde{s}} \\ t_i(\tilde{X}_i)_{\tilde{s}} \\ s_i(\tilde{X}_i)_{\tilde{t}} \\ t_i(\tilde{X}_i)_{\tilde{t}} \\ (\tilde{X}_i)_{\tilde{t}} \\ -\tilde{s}_i s_i(\tilde{X}_i)_{\tilde{s}} - \tilde{t}_i s_i(\tilde{X}_i)_{\tilde{t}} \\ -\tilde{s}_i t_i(\tilde{X}_i)_{\tilde{s}} - \tilde{t}_i t_i(\tilde{X}_i)_{\tilde{t}} \\ -\tilde{s}_i(\tilde{X}_i)_{\tilde{s}} - \tilde{t}_i(\tilde{X}_i)_{\tilde{t}} \end{pmatrix},$$

where, (26)

$$\tilde{s}_i = \frac{H_1 \cdot e\left(s_i, t_i\right)}{H_3 \cdot e\left(s_i, t_i\right)}, \tilde{t}_i = \frac{H_2 e\left(s_i, t_i\right)}{H_3 e\left(s_i, t_i\right)},$$

One should notice that the registration method may be changed to other methods, such as Kim and Lee [17].

## 3. Numerical results

A quadcopter hovering over a target with a specific FOV suffers from hovering errors. Following the above model, these errors can be exploited to provide the additional information required for scene-based NUC. A schematic configuration is presented in Fig. . Here we show how, using synthetically corrupted images and the suggested algorithm, performing image restoration and NUC of both gain and offset.

Simulated data of two space-variant study cases is used to demonstrate the proposed algorithm, a quality assessment achieved by comparing the restored image to the ground truth data.

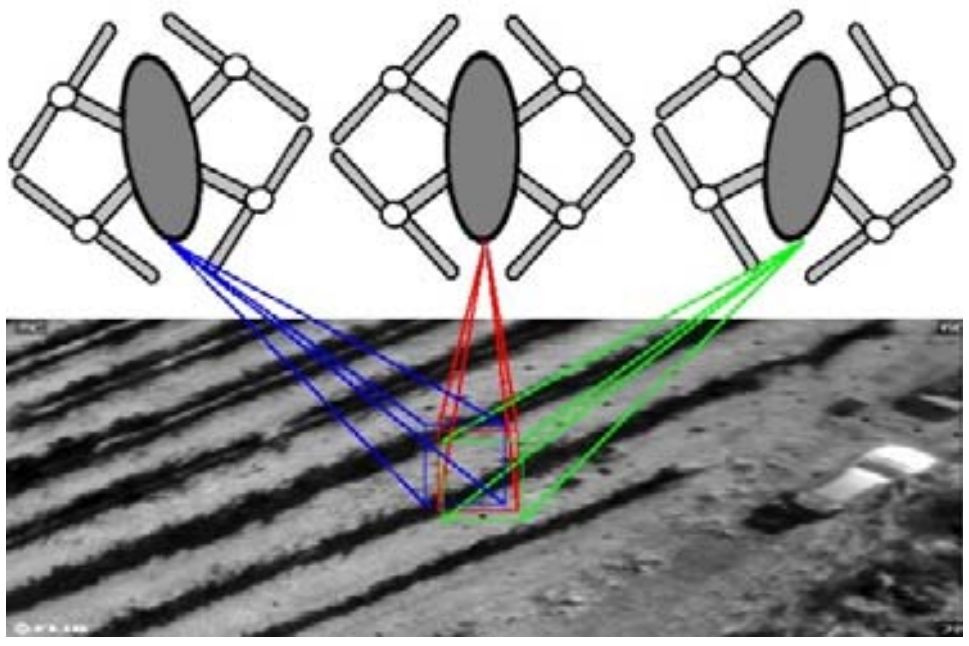

Fig. 3. Quadcopter hovering over a target with a specific FOV. Hovering error results in k slightly shifted images, each from a different projection view.

### 3.1. Simulation conditions

Ground truth images were collected in a Cabernet Sauvignon vineyard, established in Mevo-Beitar (31.729522ºN; 35.1031ºE) in the Judean hills region in Israel. The aerial surveys were performed in spring 2017 and 2018 with a FLIR A655Sc camera. The camera was mounted on a quadcopter imaged from approximately 60 meter height. Images first saved as a SEQ video converted into a .jpg images using FLIR Tools software, the images used as the ground truth data for the simulations. The temperature range was 27.1°C to 51.7°C, represented by 255 gray values. Where each gray level equals ~0.1°C.

The method was tested over 8 disjoint FOVs. For the sake of saving on computation effort, the image size was 66 × 66 pixels. For each FOV a series of 64×64 images was taken. That is, we used m = 64 images divided into 8 groups such that each group contains 8 images of the same FOV and the intersection for every pair of images from different groups is empty.

To mimic the hovering error, the images were taken while subjected to random homography transformations. Each image was further subjected to gain and offset transformations and Gaussian noise was added such that the resultant signal-to-noise ratio was 1000. For consistency, we adopt two cases studied for unknown space-variant "radial-like" and "sine-like" profiles (Eq. 27a and Eq. 27b, respectively). Which have been explored in our previous work [7]. The gain is normalized such that its mean is 1, and the offset is normalized such that its mean is 0. Functions of the unknown offsets and gains are:

$$\begin{aligned} g(s,t) &= \left((x-w)/2\right)^2 + \left((x-h)/2\right)^2 \\ d(s,t) &= 4\,std(x)\cdot\left(0.5\cdot\left((x-w)/2\right)^2 + \left((x-h)/2\right)^2\right) \end{aligned} \quad (27a)$$

$$\begin{aligned} g(s,t) &= \left(1+0.3\sin(s/2.5)\sin(t/7.5)\right) \\ d(s,t) &= (3.5\,std(x)\sin\left((s+t)/5\right)\sin\left((s-t)/10\right) \end{aligned} \quad (27b)$$

Where std(x) is the standard deviation of the objects ensemble. The profiles of the "radial" and "sine" space-variant functions are respectively presented in Fig.4 and Fig.5,.

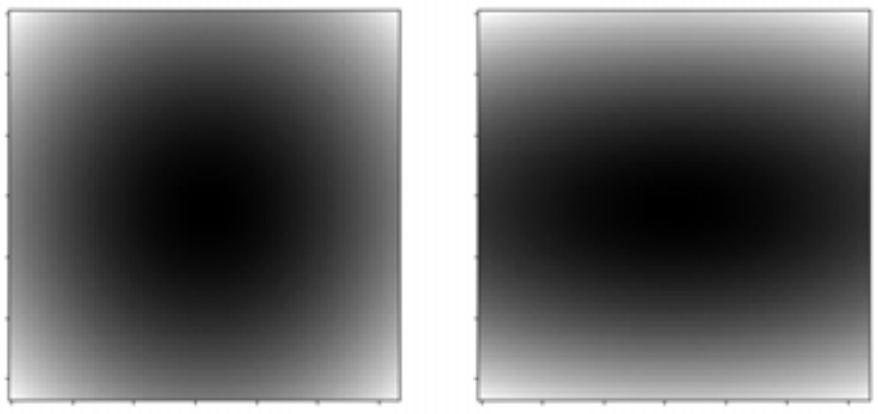

Fig. 4. Radial spatial profiles: gain (left) and offset (right).

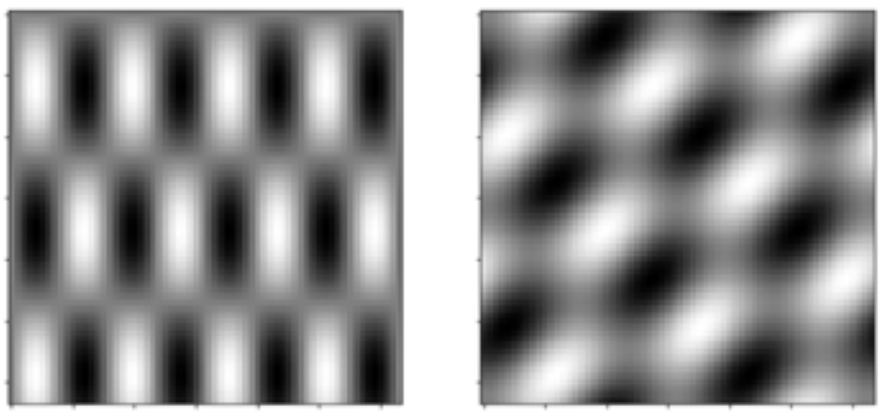

Fig. 5. Sinusoidal spatial profiles: gain (left) and offset (right).

Assuming that the camera is mounted on the UAV through a gimbal, the hovering errors can be compensated for by the gimbal mechanism with some residues. The homography transformations followed the parameterization in Raviv [18], such that: $t_z \sim Uniform(-0.5[\mathrm{m}], 0.5[\mathrm{m}])$, γ~Uniform(-5° ,5°), and α,β,ψ,ξ~Uniform(-0.05°,0.05°). where $t_x, t_y, t_z$ denote the translations in axes x, y, z, respectively, α,β,γ denote roll, pitch and yaw angles, respectively, and ψ, ξ denote the angles by which it is required to rotate axis z around axes x and y, respectively, to obtain the direction relative to the earth (inertial) system. Uncertainties for $\alpha, \beta, \psi, \xi$ of $0.1°$ are likely, due to commercial gimbals' limitations, which achieve angular inaccuracies of $0.01°$ [19]. The $t_z$ uncertainty of 0.5 is due to a vertical hovering error [20]. In addition, $t_x$ and $t_y$ are drawn such that the image's translation is no larger than 1 pixel; this can be achieved by discarding images with higher translation in the first registration step. The first image of every object $y_{i1}$ is taken without a transformation, i.e., $H_{i1}$ is considered to be the unit matrix.

### 3.2. Image restoration

The algorithm's results for the "radial" case and "sine" case are depicted alongside the ground truths in Fig. and Fig. , respectively. A series of *n* = 8 was considered, where each image covers a slightly different part of the FOV. The restoration was performed in the overlapping area. Thus, the presented error is computed without considering the image's boundary (which appears in white). The results of the "radial" case in 4 of the 8 FOVs are presented in Fig 6. Each FOV is presented on a different line. The rows, ordered from left to right, are the corrupted image (the algorithm input), the ground truth, the restored image, and in the last row, the difference map. The figures show very good restoration quality. The root mean square error (RMSE) of the restoration is marked above each restoration image and is within the range of 0.043 to 0.073 gv. This is equivalent to a seamless temperature error that is lower than 0.007 °C.

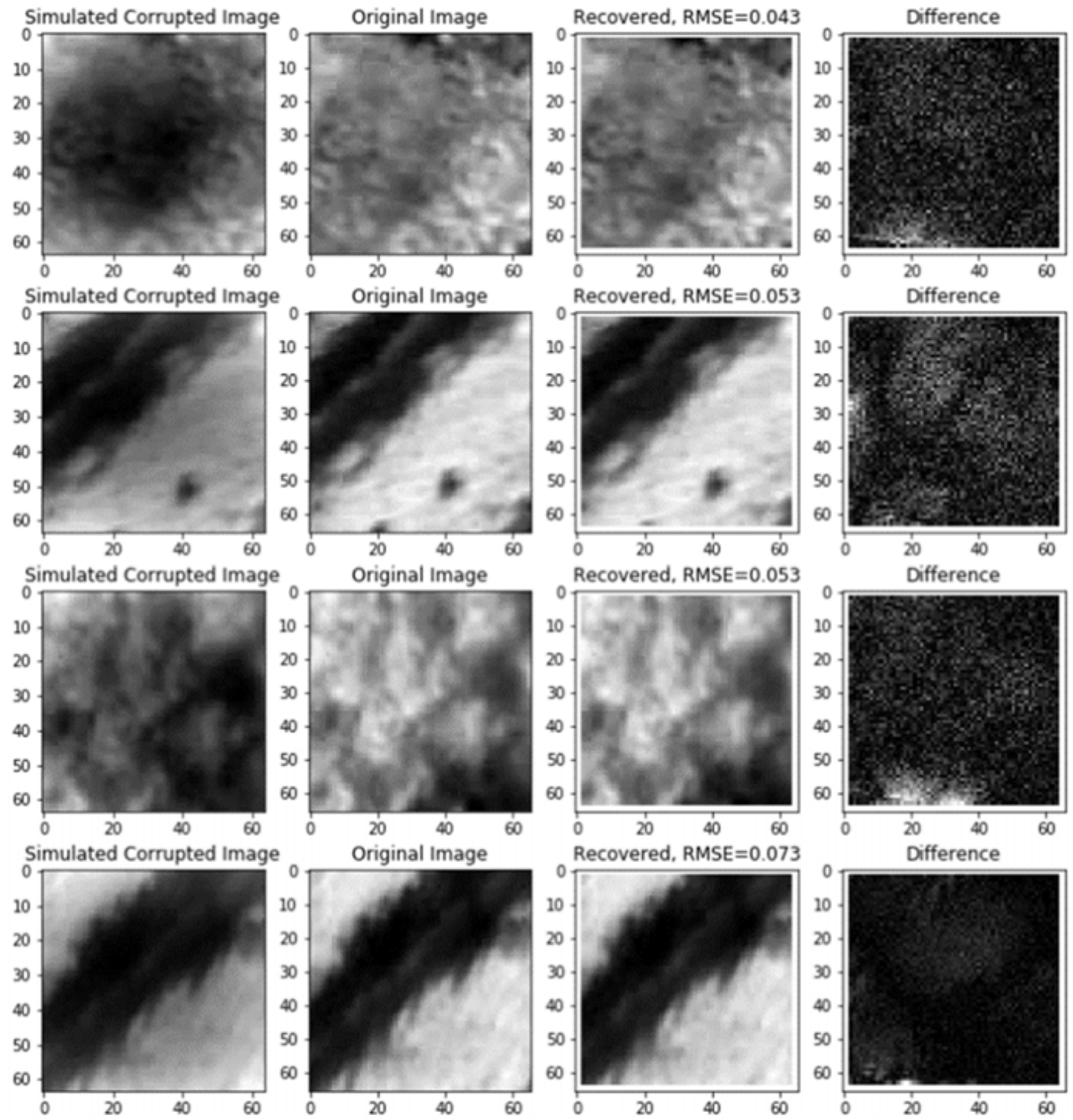

Fig. 6. Simulated results with radial profile, from left to right: images after they were subjected to gain, offset, homograpy, and noise; corrupted image; ground truth image; our algorithm's estimations; difference between ground truth images and our estimations.

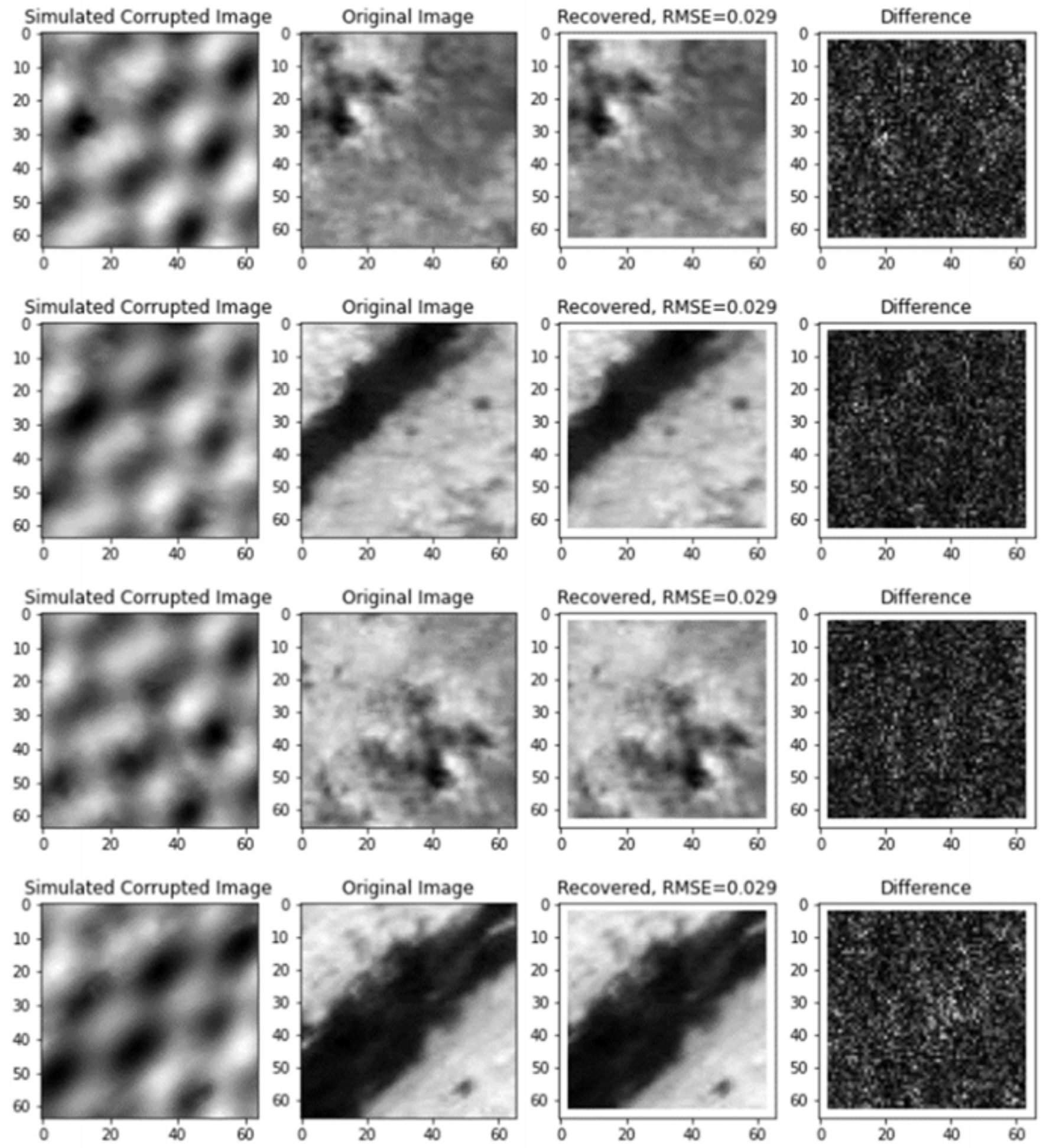

Fig. 7. Simulated results with sinusoidal profile, from left to right: images after they were subjected to gain, offset, homograpy, and noise; corrupted image; ground truth image; our algorithm's estimations; difference between ground truth images and our estimations.

The results of the "sine" case are presented in Fig. 7. For the sake of simplicity, the structure of the figure is identical to that of Fig. 6. As in the previous case, the restoration results have very good RMSE errors in the range 0.029 gv.

### 3.3. Accuracy of gain and offset recovery

The gain, offset, and series of images were jointly estimated. An example of the estimation quality of the parameters $g$ and $d$ for the radial case is shown in Fig.8. The left column presents the ground truth profiles, the middle column displays the restored profiles and the right column shows the two-dimensional distribution of the gain error. As before, the restoration is limited to the over lapping zone in the series of images. Thus, the error is computed without considering the image boundaries. The results of both gain and offset are very accurate with RMSE of 0.17% in gain and 0.059 out of 255 gv in offset.

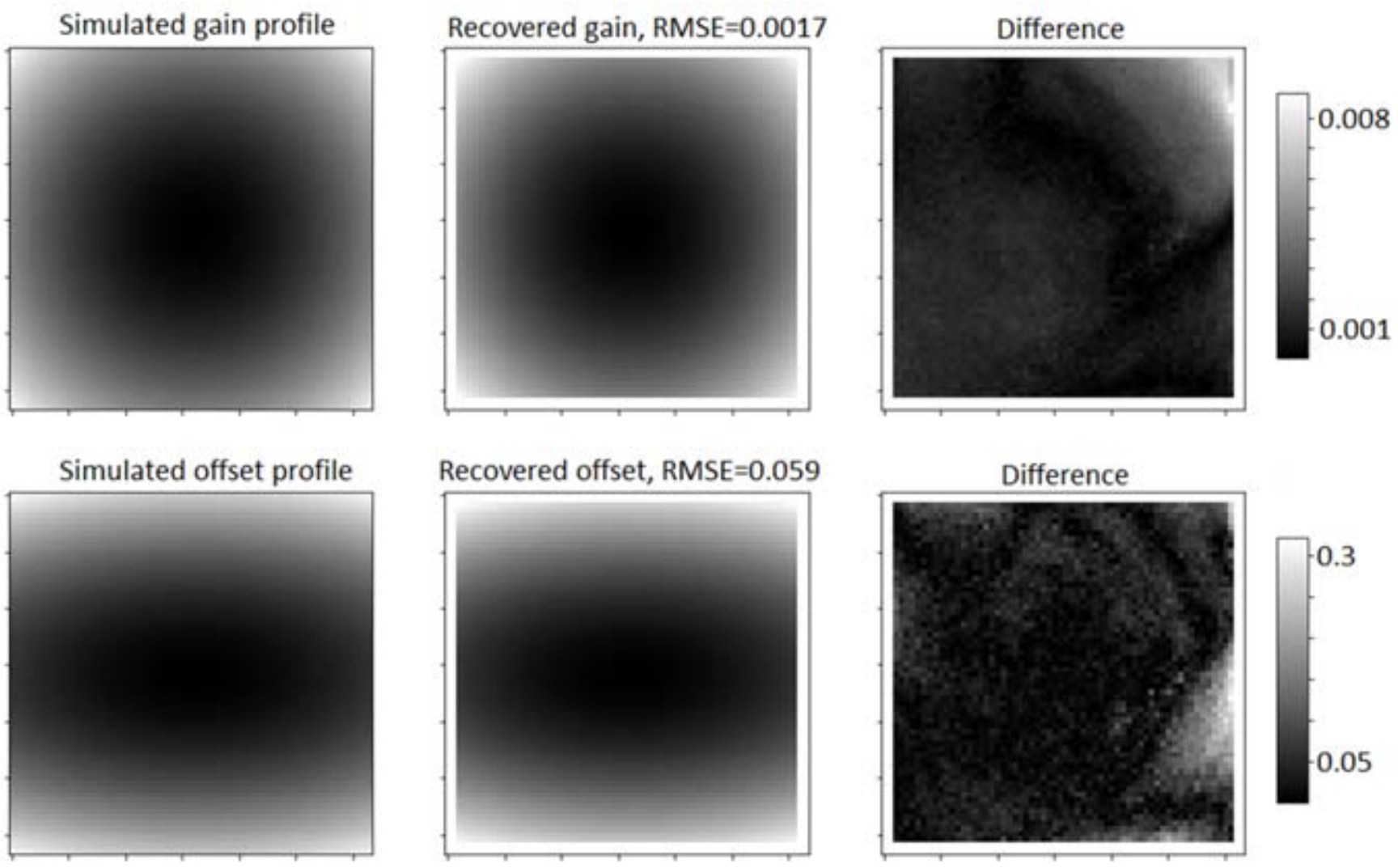

Fig. 8. Simulated gain and offset profiles, their estimations, and their (absolute value) differences. Gain values range from 0.7 to 1.3 and their estimation errors range from -0.001 to 0.001. Offset values range from -200 to 200 and their estimation errors range from -0.3 to 0.3.

### 3.4. Influence of k on the precision of the image restoration

The number of required images per FOV is an important experimental factor. Very high values may risk the assumption of permanent unknown offset and gain during the imaging process. In Table 1, we show the average Pearson coefficients and RMSE between our restoration and the ground truth over 30 different results, with *k* values of 10, 20, 30, and 40; the coefficients show an almost constant value, larger than 0.999999, which reflects a very high correlation between the algorithm's restoration result and the ground truth. The RMSE values of the experiments are presented in terms of gray levels. In all cases, the RMSE is a little over 0.027. Thus, from a practical point of view, *k* = 10 is sufficient.

Table.1. Accuracy of gain and offset recovery

| **k** | 5 | **10** | **20** | **30** | **40** |
|---|---|---|---|---|---|
| **⟨PC⟩** | 0.99999971 | 0.99999986 | 0.99999985 | 0.99999987 | 0.99999987 |
| **<RMSE>** | 0.0435 | 0.02722 | 0.0277 | 0.0275 | 0.028 |

⟨PC⟩ is the average Pearson coefficient, between algorithm estimation and ground truth. <RMSE> is the average root mean square error, between algorithm estimation and ground truth in gray levels.

### 3.5. Image restoration with the previous proposed method

In a previous work [7] it was suggested to perform the joint estimation $o(\{\hat{x}_i\}, \hat{g}, \hat{d})$, by the means of measuring series of n>2 pairs of focus and defocus images ($y_{i1}$and $y_{i2}$) of different objects $\{x_i\}$. Fig.9 presents simulations results of the previous algorithm when the pairs are composed of images with homography errors resulted from the gimbal as in the present work. The dataset and homography errors are taken from this work, where focused and defocused kernels are identical to [7].

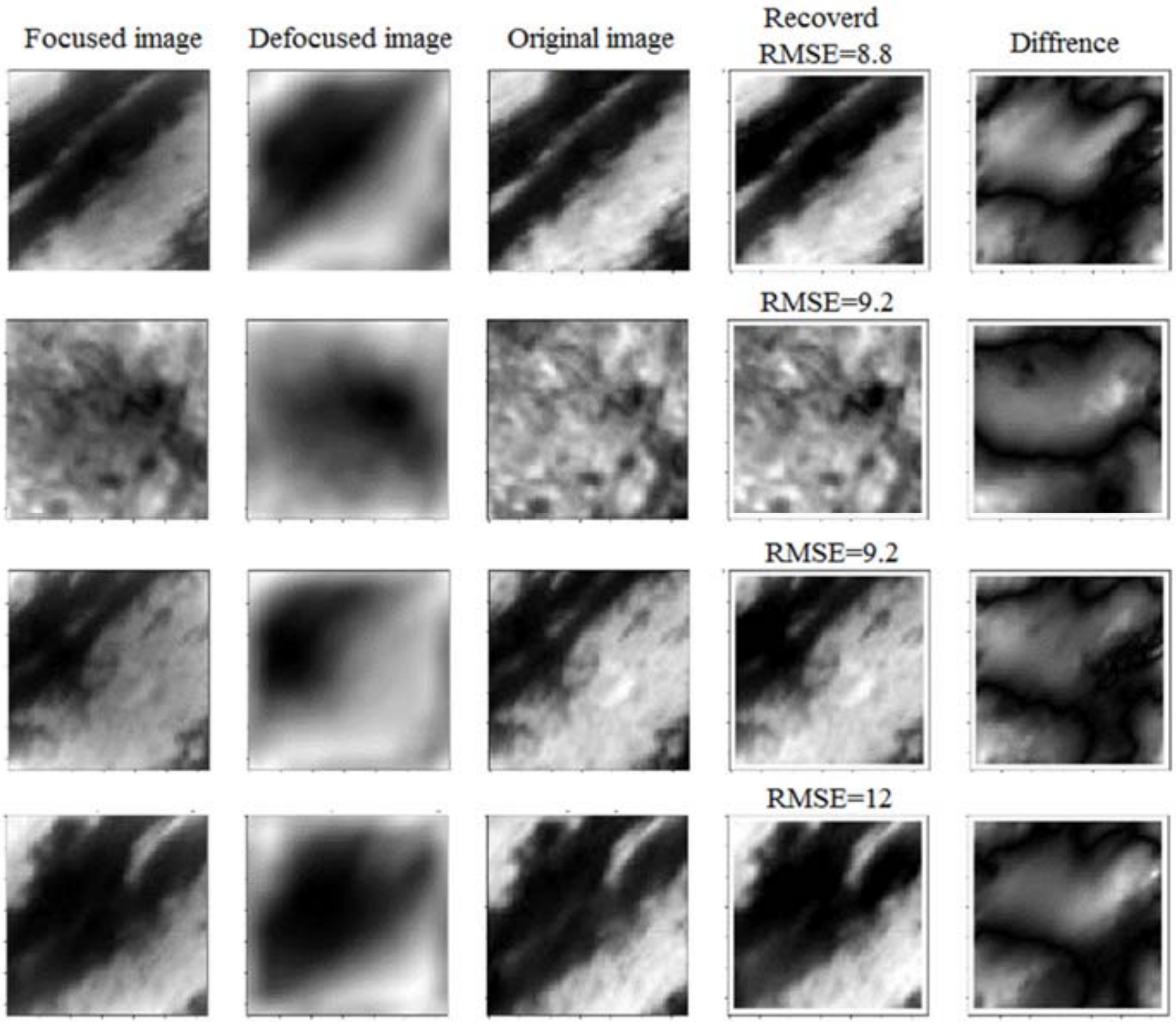

Fig. 9. Simulated results with radial profile, from left to right: images after they were subjected to gain, offset, homograpy, and noise; in focused corrupted image; defocused corrupted image, ground truth image; our algorithm's estimations [7]; difference between ground truth images and the estimations.

Observing the results average ensemble RMSE is 9.225 gv which equals to 1.1 °C which is significantly larger (two orders of magnitude) than the results in the current work. In addition, the use of the relative displacement instead of defocusing allows a simplification of the requirements from the hardware.

## 4. Summary and conclusions

The affordable UC-FPAs are not isolated or cooled. This results in gain and offset drift under environmental load. The drift in the FPA's gain and offset restricts the thermographic quality of these cameras. Thus, calibrating the camera readout is an important challenge. Previous work on scene-based and registration-based NUC solved the problem using a sequence of shifted images, performing a registration stage followed by a correction stage. These studies could only correct the offset. In addition, they used partial models of motion, which cannot properly model the realistic motion of a camera on a UAV. Our work extends the motion model to homography transformations, which allow our suggested algorithm to take into account translations, rotations, and tilts associated with the UAV's vibrations. The problem is formulated as a minimization problem. It is suggested to solve it using alternating minimization, with the registration being one of its stages. The ability to accurately correct the acquired images, corrected for both offset and gain, from a series of images, was demonstrated in simulations of an unknown "radial-like" and "sine-like" offsets and gains functions. Results showed low restoration error with a propitious RMSE (gv) that, under ideal assumptions, is equivalent to a mean restoration error of less than 0.01℃ which is very accurate. This in comparison to 1.1°C error, resulted from a previous proposed method which also requires for more complicated focusing hardware.

**Acknowledgments**

This work was supported by the Israeli Ministry of Agriculture's Kandel Program for funding this research under grant no. 20-12-0030.

**Declarations of interest:** none